\documentclass[submission,copyright,creativecommons]{eptcs}
\providecommand{\event}{WACA 2025} %

\usepackage{graphicx}
\graphicspath{}

\usepackage{amssymb}
\usepackage{xcolor}
\usepackage[noabbrev,capitalise]{cleveref}

\usepackage[inline]{enumitem}

\usepackage{array}             %
\usepackage{tabulary}          
\usepackage{booktabs}          %
\usepackage{calc}              %

\newlength{\tabwidthi}
\newlength{\tabwidthii}
\newlength{\tabwidthiii}
\newlength{\tabwidthiv}
\newlength{\tabwidthv}
\newlength{\tabwidthvi}

\newcommand{\tabsize}{\footnotesize}

\newcommand{\thead}[1]{\tabsize\textbf{#1}}

\newcommand{\mdash}[1][]{#1---#1}

\newcommand{\ie}[1][\@ ]{i.e.,#1}

\newcommand{\eg}[1][\@ ]{e.g.,#1}

\newcommand{\cf}[1][\@ ]{cf.#1}

\newcommand{\wrt}[1][\@ ]{w.r.t.#1}

\newcommand{\resp}[1][\@ ]{respectively#1}

\usepackage{hyperref}

\providecommand{\thisvolume}[1]{this volume of EPTCS, Open Publishing Association}

\title{Adaptable TeaStore\thanks{Work
    partially supported by French ANR project SmartCloud
    ANR-23-CE25-0012, by PRIN project FREEDA (CUP: I53D23003550006)
    funded by the frameworks PRIN (MUR, Italy) and Next Generation EU, by project PNRR CN HPC - SPOKE 9 - Innovation Grant LEONARDO - TASI - RTMER funded by the NextGenerationEU European initiative through the MUR,
    and by INdAM - GNCS 2024 project MARVEL, code CUP E53C23001670001.}}

\author{Simon Bliudze\institute{Univ. Lille, Inria, CNRS, Centrale Lille, UMR 9189 CRIStAL, F-59000 Lille, France}
\and
Giuseppe De Palma\institute{Alma Mater Studiorum - Università di Bologna, Bologna, Italy \\ Olas Team, INRIA, Sophia Antipolis, France}
\and
Saverio Giallorenzo\institute{Alma Mater Studiorum - Università di Bologna, Bologna, Italy \\ Olas Team, INRIA, Sophia Antipolis, France}
\and
Ivan Lanese\institute{Alma Mater Studiorum - Università di Bologna, Bologna, Italy \\ Olas Team, INRIA, Sophia Antipolis, France}
\and
Gianluigi Zavattaro\institute{Alma Mater Studiorum - Università di Bologna, Bologna, Italy \\ Olas Team, INRIA, Sophia Antipolis, France}
\and
Brice Arléon Zemtsop Ndadji\institute{Univ. Lille, CNRS, Inria, Centrale Lille, UMR 9189 CRIStAL, F-59000 Lille, France}
}

\def\authorrunning{Bliudze et al.}
\def\titlerunning{Adaptable TeaStore}

\begin{document}
\maketitle

\begin{abstract}
Modern cloud-native systems require adapting dynamically to changing operational
conditions, including service outages, traffic surges, and evolving user
requirements. While existing benchmarks provide valuable testbeds for
performance and scalability evaluation, they lack explicit support for studying
adaptation mechanisms, reconfiguration strategies, and graceful degradation.
These limitations hinder systematic research on self-adaptive architectures in
realistic cloud environments.

To cover this gap, we introduce \emph{Adaptable TeaStore}, an extension of the
renowned TeaStore architecture that incorporates adaptability as a first-class
design concern. Our extension distinguishes between mandatory and optional
services, supports multiple component versions\mdash with varying resource
requirements and functionality levels\mdash considers the outsourcing of
functionalities to external providers, and provides local cache mechanisms for
performance and resilience. These features enable the systematic exploration of
reconfiguration policies across diverse operational scenarios.

We discuss a broad catalogue of reference adaptation scenarios centred around
Adaptable TeaStore, useful to evaluate the ability of a given adaptation
technology to address conditions such as component unavailability, cyberattacks,
provider outages, benign/malicious traffic increases, and user-triggered
reconfigurations. Moreover, we present an open-source implementation of the
architecture with APIs for metrics collection and adaptation triggers, to
enable reproducible experiments.
\end{abstract}

\section{Introduction}
\label{sec:intro}

In modern cloud software architectures, the ability to adapt services to
changing conditions is increasingly essential. Distributed systems must remain
operational despite fluctuating workloads, partial failures, and evolving user
requirements. While state-of-the-art approaches often adopt the
microservice~\cite{DGLMMMS17} or serverless~\cite{JSSTKPSCKYGPSP19} paradigms,
the architectural benchmarks available today largely emphasise performance and
scalability rather than explicit support for adaptation. This gap limits the
capacity of researchers and practitioners to study reconfiguration strategies,
graceful degradation, and the substitution of external providers in a systematic
and comparable way.

TeaStore~\cite{KESBGK18} has emerged as a reference benchmark for
microservice-based systems. The benchmark provides a realistic yet tractable
environment for studying performance trade-offs, resource management, and
deployment variability. However, TeaStore's original design does not explicitly
account for adaptation, leaving out many scenarios that arise in real-world
deployments where applications must cope with service outages, degraded
performance, cyberattacks, or surges in demand. Addressing these limitations
requires a benchmark that supports both mandatory and optional services,
multiple implementation variants, and external dependencies, together with
mechanisms for fallback and reconfiguration.

To cover this gap, we propose \emph{Adaptable TeaStore}, a variant of TeaStore
that includes adaptability as a core feature. Adaptable TeaStore substantially
extends the architecture of the original TeaStore benchmark with characteristics
that support the reconfiguration of the application under varying operational
conditions. Central to this design is the distinction between mandatory
services, which guarantee baseline functionality, and optional ones, which can
be selectively enabled or disabled to trade features for resilience or
efficiency. Several services are offered in multiple ``flavors'', ranging from
lightweight to fully featured implementations, thereby supporting fine-grained
adaptation to resource availability and quality-of-service constraints.
Moreover, functionalities, such as data persistence, may be outsourced to
external providers, reflecting modern cloud deployment practices, while local
cache mechanisms and static fallback solutions ensure continuity in the presence
of provider failures or disconnections. Taken together, these characteristics
make Adaptable TeaStore a flexible benchmark for studying reconfiguration
strategies, graceful degradation, and provider substitution within a realistic
microservice setting. 

Looking at the state of the art, Adaptable TeaStore acts as a bridge between the
landscape of microservice and cloud benchmarks and that of the
adaptation-centric ones, complementing existing microservice-style
performance-oriented benchmarks such as Acme Air~\cite{acmeair}, Sock
Shop~\cite{sockshop}, and DeathStarBench~\cite{GZCSRKBHRJHPH19} and
adaptation-oriented ones, like mRUBiS~\cite{V18}.

The paper is structured as follows. We begin, in \cref{sec:teastore}, by
recalling the original TeaStore architecture, emphasising its strengths as a
research-grade reference system and its limitations in the context of adaptive
behaviour. Moving to \cref{sec:adaptable_teastore}, we present our extensions,
which enrich TeaStore's architecture with configurable and outsourceable
services, service flavours offering different levels of functionality, and local
cache mechanisms that ensure graceful degradation when external dependencies
fail. Building on these extensions, we discuss several configuration levels that
capture different operating modes, from minimal barebone deployments to
full-featured configurations relying on external providers.
To demonstrate the range of adaptation situations Adaptable TeaStore can
support, we present a catalogue of adaptation scenarios that stress a given
adaptation technology's ability to manage diverse operational conditions. These
scenarios include coping with unavailable data sources, mitigating cyberattacks
on external services, handling benign or malicious traffic surges, and
responding to user-triggered reconfigurations. Together, these scenarios provide
a systematic basis for evaluating adaptive mechanisms, resilience strategies,
and their trade-offs. 
As a second contribution, in \cref{sec:implementation}, we present an
implementation of Adaptable TeaStore that we provide as open source code and as
images on Docker Hub to be used as an experimentation platform for the
adaptation of microservice-based applications.  We describe the interfaces
provided for querying metrics and for triggering adaptation actions.
In \cref{sec:related_work}, we position our contribution \wrt related work. We
conclude, in \cref{sec:conclusion}, by drawing closing remarks on the broader
implications of Adaptable TeaStore and outlining community-contributed
directions for future extensions.

\section{TeaStore}
\label{sec:teastore}
TeaStore~\cite{KESBGK18} is a well-known reference benchmark for microservice-based systems. It involves 5 different services: WebUI, Auth, Persistence, Image
Provider, and Recommender.

\begin{figure}[t]
      \centering
      \includegraphics[width=0.8\textwidth]{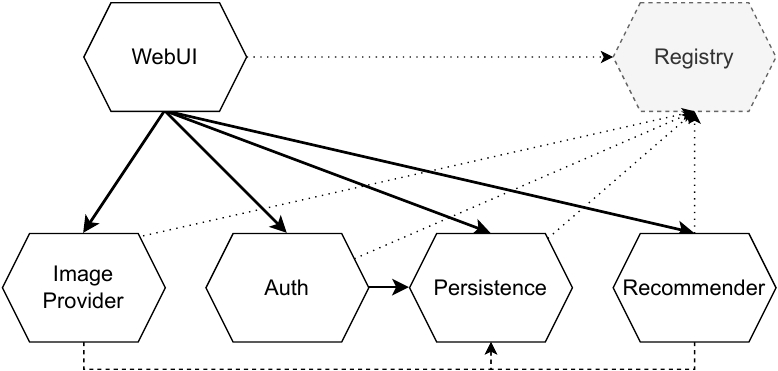}
      \caption{TeaStore Architecture, adapted from von Kistowski et al.~\cite{KESBGK18}}
      \label{fig:teastore}
\end{figure}

In \cref{fig:teastore}, we show an overview of the TeaStore architecture. The
WebUI service is the entry point for the user, and it is the main service that
interacts with all other services. The Persistence service is the layer on top
of the database. It is used by the WebUI to retrieve and store data and by the
Authentication service to retrieve user data, which is then passed to the WebUI.
On startup, the Persistence service populates the database with data. The Image
provider and Recommender both connect to a provided interface at the Persistence
service. However, this is only necessary on startup (dashed lines). The Image
provider must generate an image for each product, whereas the Recommender needs
the current order history as training data. Once running, only the
Authentication and the WebUI access, modify, and create data using the
Persistence. The Registry service will be described in \cref{sec:implementation}.

We conclude this section by summarising the description of TeaStore's components
from the benchmark's original proposal~\cite{KESBGK18}.

\paragraph{The WebUI service} manages the user interface and contacts all the
other services to retrieve and display data. It compiles and serves Java
Server Pages (JSPs) with the categories, products, recommendations, and
images. It performs preliminary validation on user inputs before sending them
to the Persistence service.

\paragraph{The Image Provider service}
handles the serving and resizing of images in various sizes
for the WebUI. It uses a cache to optimise performance, resizing and caching
images if necessary. The system employs a least-frequently-used caching strategy
to reduce resource demand, and response time depends on whether an image is
already cached or needs resizing.

\paragraph{The Authentication service}
verifies user login and session data, using BCrypt for login
verification and SHA-512 hashes for session validation. Session data include
shopping cart content, login status, and order history. The performance depends
on the volume of session data. The service remains stateless since all session
data is passed to the client.

\paragraph{The Recommender service}
provides product recommendations based on items other
customers bought, on the products in a user's current shopping cart, and on the
product the user is viewing at the time. It maintains coherence between
different instances by sharing training data (additional Recommender instances
query existing instances for their training data-set). Several recommendation
algorithms can be used, including two `Slope One' variants and an order-based
nearest-neighbourhood method, with different algorithms optimising for either
memory or CPU performance. There is a fallback algorithm based on overall item
popularity.

\paragraph{The Persistence service}
manages access and caching for the store's relational
database, which stores data on products, categories, purchases, and users. There
is an internal cache to improve scalability and reduce the load on the database.
The cache is kept coherent across multiple Persistence service instances. All
data inside the database itself is generated at the first start of the initial
persistence instance.

\section{Adaptable TeaStore}
\label{sec:adaptable_teastore}

The original TeaStore architecture is static, lacking scenarios where the
structure of the system needs to change due to failures and variations of
environmental conditions or user requirements. To enhance the case study in this
direction, we extend it with optional services and external dependencies. The
latter kind are functionalities that the entity deploying the system to provide
a service to clients, henceforth called the \emph{company}, has no control over.

\begin{figure}[t]
    \centering
   \includegraphics[width=\textwidth]{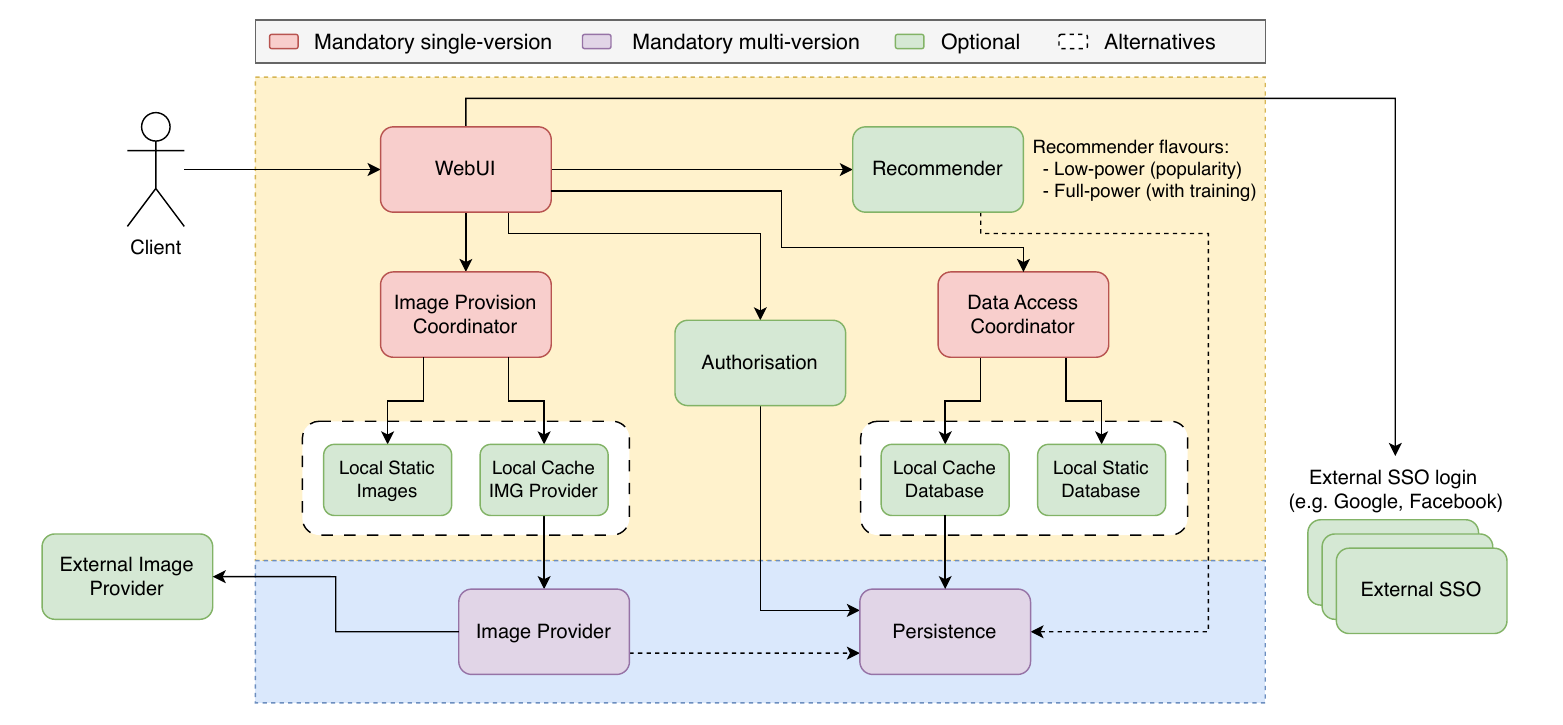}
    \caption{Adaptable TeaStore Architecture (functionalities shown within the coloured areas are internal, those in the blue area are outsourceable, in the yellow area\mdash non-outsourceable (\cf \cref{tab:func_summary}))}
    \label{fig:teastore-variant}
\end{figure}

The Adaptable TeaStore architecture is shown in \cref{fig:teastore-variant}.
Below, we present our extensions to the original architecture and the proposed
adaptation scenarios.

We first pinpoint some terms that we use in the description of the extension of
the TeaStore architecture. 
Functionalities of an architecture are either \emph{mandatory} or
\emph{optional}. Moreover, functionalities of a given architecture have
different \emph{provision} modalities: either by components managed by the
company or by third parties. In the first case, we classify the functionality as
offered by an \emph{internal} component. Internal components can be either
hosted on premises or deployed on Cloud. Otherwise, a third party offers that
functionality as an \emph{external} service. Some functionalities can be
provided either through internal components or external services, and we call
them \emph{outsourceable}. Another dimension is that of \emph{service flavours},
\ie alternative implementations of a component that provides a given
functionality and that one can deploy interchangeably.

Since (Adaptable) TeaStore adopts a microservice architecture, internal
functionalities are provided by (micro)services. Therefore, below, we use the
terms service and component interchangeably.

\subsection{Extensions of the TeaStore Architecture}
\label{sec:extensions}

We categorise the functionalities of the Adaptable TeaStore as follows.

\paragraph{Mandatory and Optional Functionalities} The WebUI, Image Provider,
and Persistence services provide functionalities which are mandatory
for the system to operate. The functionalities provided by the Recommender and
by the Authentication service are both optional.
Moreover, we introduce in Adaptable TeaStore the optional functionality of
Single Sign-On (SSO) authentication. Note that the two authentication techniques
are not mutually exclusive. Indeed, while the two functionalities provide the
same outcome (authenticating users) their modalities significantly change: the
one provided by the Authentication service involves two parties, a client and a
server, where the latter has the necessary information to authenticate the user,
while SSO involves three parties, a client, a relying party, and an identity
provider, where the relying party trusts the identity provider to give access to
clients. An Adaptable TeaStore instance can work without
either of the functionalities, allowing users to anonymously browse products,
but it can also work with either or both functionalities available, since they
manage the authentication of different sets of users.

\paragraph{Outsourceable Functionalities} The functionalities provided by the
Image Provider and Persistence services can come either from an
external provider, \eg Database-as-a-Service (DaaS) platforms for the
Persistence functionalities (\eg Firebase, Supabase), or from the respective
homonymous internal services\mdash this scenario includes also multi-tenant
deployments where multiple applications of the same company share the same
services.
    
\paragraph{External Functionalities} As mentioned above, we introduce the
possibility for users to authenticate through Single Sign-On as a coexisting
alternative to the internal Authorisation service.  This option reflects typical applications that integrate third-party login options (\eg Google,
Facebook, etc.).

\paragraph{Service Flavours} In Adaptable TeaStore there are three components,
each coming in two flavours. The Image Provider comes either in its
full-featured flavour, named after the service itself, which can dynamically
resize images according to request parameters (\eg for smaller or larger
screens) or its lightweight alternative, which provides only static images,
called \emph{Local Static Images} service. The Recommender service has either a
\emph{Full-power} flavour, which runs a resource-intensive recommendation
algorithm, or a \emph{Low-power} one, which requires less resources but might
provide less accurate recommendations\mdash the latter corresponds to the fallback
algorithm based on overall item popularity. The Persistence service also has two
flavours: the one from the original TeaStore, named after itself, and a static
version that does not support writing new data (\eg it allows users to browse
products but not to buy them), called \emph{Local Static DB}.

\paragraph{Optional Local Cache Services} We further extend the original
TeaStore architecture by introducing \emph{local} (internal) \emph{cache
services}, \eg to buffer interactions with outsourced or external services.
Performance-wise, these services can reduce latency and minimise redundant
requests. Functionality-wise, the cache services can reply using cached
information when the corresponding outsourced/external functionality is not
available. We consider two such optional cache services, one for the Image
Provider and one for the Persistence service, respectively called Local Cache
Image and Local Cache DB.

\paragraph{Coordinators} To avoid modifying the WebUI code to select, \eg between Local Static Images and Local Cache Image Provider (similarly for database access), we introduce two new coordinator services\mdash \emph{Image Provision Coordinator} and \emph{Data Access Coordinator}.  The role of these services is to channel the WebUI requests according to the current configuration.  In the absence of optional local cache services, the Image Provision Coordinator (\resp the Data Access Coordinator) interacts directly with the Image Provider (\resp Persistence).

\medskip
An overview of the characteristics of the functionalities and
components of the Adaptable TeaStore architecture, we summarise them in
\cref{tab:func_summary}.

\begin{table}[t]
  \centering
  \caption{Summary of the characteristics of the functionalities/components found in Adaptable TeaStore (a bullet $\bullet$ indicates that the
    element has the given property, while a circle $\circ$ denotes its
    absence)}
  \label{tab:func_summary}
  \tabsize
  \settowidth{\tabwidthi}{Image Provision Coordinator}
  \settowidth{\tabwidthii}{\thead{Mandatory}}
  \settowidth{\tabwidthiii}{\thead{Provision Modality}}
  \settowidth{\tabwidthiv}{\thead{Outsourceable}}
  \settowidth{\tabwidthv}{\thead{\# Flavours}}
  \settowidth{\tabwidthvi}{\thead{Local Cache}}
  \begin{tabular}{@{}
      >{\raggedright}p{\tabwidthi}
      >{\centering}p{\tabwidthii}
      >{\centering}p{\tabwidthiii}
      >{\centering}p{\tabwidthiv}
      >{\centering}p{\tabwidthv}
      >{\centering}p{\tabwidthvi}
      @{}
    }
    \toprule
    \thead{Functionality}
    &
    \thead{Mandatory}
    &
    \thead{Provision Modality}
    &
    \thead{Outsourceable}
    &
    \thead{\# Flavours}
    &
    \thead{Local Cache}
    \tabularnewline\toprule
    WebUI & $\bullet$ & Internal & $\circ$ & 1 & $\circ$
    \tabularnewline\midrule
    Image Provision Coordinator & $\bullet$ & Internal & $\circ$ & 1 & $\circ$
    \tabularnewline\midrule
    Data Access Coordinator & $\bullet$ & Internal & $\circ$ & 1 & $\circ$
    \tabularnewline\midrule
    Persistence & $\bullet$ & Internal & $\bullet$ & 2 & $\bullet$
    \tabularnewline\midrule
    Image Provider & $\bullet$ & Internal & $\bullet$ & 2 & $\bullet$
    \tabularnewline\midrule
    Recommender & $\circ$ & Internal & $\circ$ & 2 & $\circ$
    \tabularnewline\midrule
    Authorisation & $\circ$ & Internal & $\circ$ & 1 & $\circ$
    \tabularnewline\midrule
    SSO & $\circ$ & External & - & - & $\circ$
    \tabularnewline\midrule
    External Image Provider & $\circ$ & External & - & - & $\circ$
    \tabularnewline\bottomrule
  \end{tabular}
\end{table}

\paragraph{Configuration Levels}

Since the above dimensions allow one to identify multiple Adaptable TeaStore
instance configurations, we fix three reference configuration levels.

\begin{enumerate}
    \item \textbf{Barebone} contains only the mandatory services, to offer
    minimum functionality for anonymous users, in their low-level flavours, if
    any, \ie the WebUI, the Local Static Images flavour of Image Provider, and
    the Local Static DB of Persistence.

    \item \textbf{Barebone + Recommender} adds, along the services of the
    Barebone configuration, the Recommender using its Low-Power flavour.

    \item \textbf{Full} contains Persistence and Image Provider in their
    high-level flavours, possibly deployed in another region than the rest of
    the other components (these services could be used in other applications
    from the same company). As an alternative, these functionalities can be
    outsourced from an external provider (\eg Supabase). The configuration also
    includes the Authorisation service, the Recommender, which can be used in
    its Full-Power flavour, and an external SSO functionality. Either when
    outsourced or deployed in a different region, the Local Cache Image and
    Local Cache DB are used to buffer the requests to the related services. Note
    that, if both the SSO functionality and the Authorisation service are not
    available, the Recommender Full-Power flavour cannot be used due to missing
    user data.
\end{enumerate}

\subsection{Adaptation Scenarios}
\label{sec:scenarios}

We now move to present a set of scenarios designed to evaluate a system's
ability to adapt across various operational challenges. Each scenario tests
specific capabilities of a given technology for managing adaptation, from
handling infrastructure failures to responding to security threats and handling
resource constraints. The scenarios cover different basic aspects. They can be
combined to form sophisticated multi-cause scenarios for deeper evaluation of
adaptation capabilities of the system.

\subsubsection{Database Unavailable}

The system is deployed in a barebone configuration with only local services. The
queries to the local database start to timeout and the WebUI becomes
unresponsive. The WebUI adapts to avoid querying the database and show a
maintenance message while the system restarts the database service. Once the
database is back online the WebUI resumes normal operation.

\subsubsection{Cyberattack on External Providers}
The system is deployed in the full configuration with an external provider for
the Image Provider and Persistence services as well as authorisation through
SSO. The external providers detect an attack (\eg a privilege escalation) and
have to take down and restart the provided functionalities. In response to this
unavailability, due to a security threat, the system enables the Local Static
Images and the Local Static DB services. The WebUI adapts by disabling the
authentication functionality, such as new logins and registrations. At some
point, the deployed external functionalities are back online and the system
returns to the full configuration, as before the attack.

\subsubsection{Cloud Provider Outage}
The system is deployed in the full configuration. The services running within
remote regions go offline due to outages (\eg energy failure). The requests from
the WebUI to these services start to timeout. The system adapts by first
switching the configuration to barebone, to be able to provide the essential
functionalities, while the unavailable services are being redeployed at a
different provider. After the redeployed services are online, the system
switches back to the full configuration.

As an example, the system is up and running with an instance of the Image
Provider deployed in a different region than the rest of the architecture. The
WebUI fetches images from the Image Provider via the Local Cache Image provider.
A user accesses a product page and the WebUI tries to fetch the image from the
local cache service. The local cache service does not have the image, it tries
to fetch the image from the remote Image Provider, but the request times out.
The system adapts the configuration by switching to the Local Static Images
service. Meanwhile, the system deploys the remote service on a different
provider and switches back to the full configuration when the service is online.

\subsubsection{Sudden Traffic Increase}

We propose the following three scenarios distinguishing among the benign and
malicious causes for the traffic increase.

\paragraph{Benign Traffic Increase}
Incoming traffic towards the WebUI increases significantly due to a genuine
increase in users. Since the traffic increase is benign, the system must adapt
by scaling out the services to handle the increased load or taking other
measures to the same purpose (\eg switch to low-power versions of the services).

As an example, on a system with the Recommender service in Full-power mode, the
Recommender service is under heavy load due to the sudden increase in user
requests. The service quality degrades because the response time increases
significantly (defined by a user-specified QoS threshold). The system adapts the
configuration by switching the Recommender service to Low-power mode, which uses
the fallback algorithm to provide recommendations based on item popularity.

\paragraph{Malicious Traffic Increase}
Incoming traffic towards the WebUI increases significantly due to a DDoS attack.
The adaptations include deploying circuit breakers between services to prevent
cascading failures, switch Authentication service to a more restrictive mode
with additional verification, the Recommender switching to Low-power mode, and
Local Cache services activating to reduce remote/external dependencies. 

\paragraph{Conditional Handling of Traffic Increase}
Incoming traffic towards WebUI increases significantly. However, no explicit
information is available about the reason for that: the increase can be either
due to a genuine increase in the number of users or to a DDoS attack. The system
must evaluate the situation and adapt accordingly. Notice that the decision and
the adaptation need not be a mutually exclusive choice between the benign and
malicious scenarios above, it can be a combination of the two.

\subsubsection{DevOps Requirements Change}

The system should be able to adapt to reconfiguration requests from DevOps. For
instance, DevOps can decide to scale out a service or switch between different
versions, thus increasing/decreasing the amount of resources available to it or
modifying the features of the architecture (\eg removing or adding
services). This event could involve supporting additional external SSO login
options or forcing a specific service (\eg Recommender) flavour. Notice that the
reconfiguration request might be incomplete, providing information only about
the required changes but not necessarily about all their dependencies. The
system should adapt by moving to a valid configuration, ideally without
downtime.

For example, the administrator of the system requires switching between local
and remote providers (\eg moving the Persistence service from an on-premises to
a on cloud location) with little to no downtime by supporting the smooth
transition with local caching as a transition layer.

\section{Experimentation Platform}
\label{sec:implementation}

\begin{table}[t]
  \caption{REST API for metrics collection (HTTP method: GET)}
  \label{tab:metrics-endpoints}
  \tabsize
  \centering    
  \settowidth{\tabwidthi}{Database response time}
  \settowidth{\tabwidthii}{/metrics/db/responseTime}
  \settowidth{\tabwidthiii}{Core TeaStore services + Persistence}
  \begin{tabular}{@{}
      >{\raggedright}p{\tabwidthi}
      >{\raggedright}p{\tabwidthii}
      >{\raggedright}p{\tabwidthiii}
      @{}
    }
    \toprule
    \thead{Metric}
    &
    \thead{Endpoint}
    &
    \thead{Providing service}
    \tabularnewline\toprule
    CPU Usage
    &
    /metrics/cpu
    &
    Core TeaStore services + Persistence
    \tabularnewline\midrule
    Memory Usage
    &
    /metrics/memory
    &
    Core TeaStore services + Persistence
    \tabularnewline\midrule
    Request details
    &
    /metrics/requests
    &
    Core TeaStore services + Persistence
    \tabularnewline\midrule
    Service status
    &
    /metrics/status
    &
    Core TeaStore services + Persistence
    \tabularnewline\midrule
    Service state
    &
    /metrics/state
    &
    Core TeaStore services + Persistence
    \tabularnewline\midrule
    Database status
    &
    /metrics/db
    &
    Persistence
    \tabularnewline\midrule
    Database response time
    &
    /metrics/db/responseTime
    &
    Persistence
    \tabularnewline\bottomrule
  \end{tabular}
\end{table}

\begin{table}[t]
  \caption{Metrics and adaptation actions exposed by the implemented Adaptable TeaStore services}
  \label{tab:extensions}
  \tabsize
  \centering    
  \settowidth{\tabwidthi}{\tabsize Image Provider}
  \setlength{\tabwidthii}{\textwidth-\tabwidthi-2\tabcolsep}
  \begin{tabular}{@{}
      >{\raggedright}p{\tabwidthi}
      >{\raggedright}p{\tabwidthii}
      @{}
    }
    \toprule
    \thead{Service} & \thead{Adaptation actions}
    \tabularnewline\toprule
    WebUI
    &
    OpenCircuitBreaker%
    ,
    CloseCircuitBreaker%
    ,
    DDoSAttackEventBroadcast%
    ,
    EnableMaintenanceMode%
    ,
    DisableMaintenanceMode%
    \tabularnewline\midrule
    Recommender
    &
    OpenCircuitBreaker%
    ,
    CloseCircuitBreaker%
    ,
    HighPerformanceMode%
    ,
    LowPowerMode%
    ,
    NormalMode%
    \tabularnewline\midrule
    Image Provider
    &
    OpenCircuitBreaker%
    ,
    CloseCircuitBreaker%
    ,
    DisableExternalImageProvider%
    ,
    EnableExternalImageProvider%
    \tabularnewline\midrule
    Persistence
    &
    OpenCircuitBreaker%
    ,
    CloseCircuitBreaker%
    ,
    DatabaseAvailableEventBroadcast%
    ,
    DatabaseUnavailableEventBroadcast%
    ,
    EnableCache%
    ,
    DisableCache%
    \tabularnewline\midrule
    Authentication
    &
    OpenCircuitBreaker%
    ,
    CloseCircuitBreaker%
    \tabularnewline\midrule
    Registry
    &
    OpenCircuitBreaker%
    ,
    CloseCircuitBreaker%
    \tabularnewline\bottomrule
  \end{tabular}
\end{table}

\begin{table}[t]
  \caption{REST API for adaptation action execution (HTTP method: POST)}
  \label{tab:actions-endpoints}
  \tabsize
  \centering    
  \settowidth{\tabwidthi}{Execute multiple actions}
  \settowidth{\tabwidthii}{/adapt/single}
  \settowidth{\tabwidthiii}{Query parameter: \texttt{actionName} \mdash[\@ ] The name of the action to execute.}
  \begin{tabular}{@{}
      >{\raggedright}p{\tabwidthi}
      >{\raggedright}p{\tabwidthii}
      >{\raggedright}p{\tabwidthiii}
      @{}
    }
    \toprule
    \thead{Action}
    &
    \thead{Endpoint}
    &
    \thead{Request parameters}
    \tabularnewline\toprule
    Execute a single action
    &
    /adapt/single
    &
    Query parameter: \texttt{actionName} \mdash[\@ ] The name of the action to execute.\newline Example: \texttt{.../adapt/single?actionName=OpenCircuitBreaker}
    \tabularnewline\midrule
    Execute multiple actions
    &
    /adapt
    &
    Body (JSON): A list of action names to be executed.\newline Example: \texttt{["OpenCircuitBreaker", "EnableCache"]}
    \tabularnewline\bottomrule
  \end{tabular}
\end{table}

We have extended the original TeaStore implementation~\cite{TeaStore-repo} 
towards providing an implementation of the Adaptable TeaStore specification from \cref{sec:adaptable_teastore}.\footnote{%
The implementation code is accessible from a \href{https://gitlab.inria.fr/adaptable-teastore/experimentation-platform}{public repository}.  Docker images are provided on \href{https://hub.docker.com/u/cerberus237}{Docker Hub}.
} The goal is to provide a reference implementation for experimental validation of tools and techniques for managing the adaptation in microservice-based applications.

Adaptable TeaStore relies on the same architecture as the original TeaStore: in
addition to the core services\mdash WebUI, Recommender, Image Provider,
Persistence, and Authentication\mdash discussed in the previous sections, it
comprises a Registry service (see \cref{fig:teastore}) and a service running a
MariaDB database (not shown in the figure). We have modified the implementations
of the core TeaStore services and of the Registry service to allow collection of
metrics and triggering the adaptation actions.\footnote{%
Coordinators, local cache services and static flavours proposed in \cref{sec:extensions} will be implemented in future work.
}

Provided metrics are summarised in \cref{tab:metrics-endpoints}. All the core
services provide the CPU usage, memory usage, request count, service state, and
service status.  The service state is always \texttt{RUNNING} for all services
except Recommender and Image Provider. The Recommender service has an additional
state \texttt{TRAINING\_DATA}, whereas Image Provider has the additional state
\texttt{GENERATING\_IMAGES}. The \texttt{status} request returns the last
heartbeat date and time, and the hosting server id. Additionally, the
Persistence service provides the database status and the database response time.
The database status report comprises the database response time, network status,
and the numbers of active connections and pending queries.  It should be noted
that any of the above requests can be used to determine whether a service is
available or not and to measure its response time. 

Provided adaptation actions are summarised in \cref{tab:extensions}.  All
services implement the circuit breaker functionality.  The three event broadcast
actions allow sending push notifications to selected services upon the
corresponding events: WebUI can notify Persistence, Recommender, Image provider,
and Authentication about DDoS attacks; Persistence can notify WebUI and
Recommender about database availability changes. Other actions reflect the
specifications in \cref{sec:adaptable_teastore}.\footnote{%
The external image provider is currently implemented by querying
\url{https://ui-avatars.com}.
}

Both the metrics and the adaptation actions can be accessed through a RESTful
API, via corresponding \texttt{GET} and \texttt{POST} actions. For the  metrics
collection, each service in the Adaptable TeaStore exposes a REST API with a
list of available endpoints summarised in \cref{tab:metrics-endpoints}. For the
adaptation actions, each service exposes two endpoints (see
\cref{tab:actions-endpoints}). The \texttt{/adapt/single} endpoint allows a
single adaptation action to be executed by passing its name. The \texttt{/adapt}
endpoint allows several adaptation actions to be executed sequentially by
passing the list of their names.

The usage of the metrics and adaptation actions discussed above is illustrated
by \texttt{AdaptiFlow}~\cite{ZBQ25}, a framework that provides an abstraction
layer for microservice-based applications focusing on the Monitor and Execute
phases of the MAPE-K loop.

\section{Related Work}
\label{sec:related_work}

Several benchmarks and reference architectures have been proposed over the last
decade to evaluate cloud-native and distributed systems, each with distinct
traits and emphases. 

Acme Air~\cite{acmeair} is one of the earliest microservice-based benchmarks,
originally designed as a sample web application with Java and Node.js
implementations, and later adopted for studies of autoscaling and cloud
elasticity~\cite{UNO16,GCFMIS17}. The benchmark primarily highlights performance
evaluation and scaling of services, but it does not incorporate explicit
adaptation strategies beyond autoscaling policies.

Sock Shop~\cite{sockshop} was subsequently introduced as a demonstration system
for microservice tooling, resilience patterns and observability practices. While
Sock Shop provides a compact, retail-oriented application that has become widely
used for tutorials and chaos engineering exercises, it is not designed with
adaptation aspects such as feature toggling or graceful degradation.

TrainTicket~\cite{ZPXSXJZ18} represents a further step in realism, providing a
complete ticket booking system that emphasises complexity, distributed
transaction management and large-scale testing, designed for evaluating
engineering practices and system integration rather than adaptation.

As mentioned, TeaStore~\cite{KESBGK18} is a research-grade reference
application, explicitly designed to facilitate benchmarking, modelling and
resource management studies, with carefully controlled variability in
performance and deployment options. TeaStore's original definition does not
specify adaptation scenarios, which we
provide
in this proposal.

DeathStarBench~\cite{GZCSRKBHRJHPH19} broadened the scope of benchmark
applications by introducing a suite of end-to-end cloud architectures spanning
social networks, media services, e-commerce, and banking. The suite enables the
study of fan-out patterns, QoS, scalability and hardware-software co-design
implications. While diverse and realistic, DeathStarBench's benchmarks are fixed
in functionality and offer no adaptation scenarios like functional degradation
and provider variability.

Adaptable TeaStore positions itself within this context by combining the realism
and familiarity of an established microservice benchmark architecture (TeaStore)
with explicit, targeted adaptation scenarios. Like Acme Air, Sock Shop,
TrainTicket and DeathStarBench, Adaptable TeaStore targets performance and
scalability experiments, but it goes beyond these aspects by evaluating
reconfiguration policies, graceful degradation, and provider substitution.

Leaving microservices, proposals worth mentioning are mRUBiS~\cite{V18}, SPEC
Cloud~\cite{BSW17},  FaaSdom~\cite{MFKS20}, BeFaaS~\cite{GPBSZB21}, and SeBS~\cite{CKBPH21}.

mRUBiS is a component-based e-commerce system explicitly built to evaluate
self-adaptation. mRUBiS targets self-healing and self-optimisation techniques
and makes adaptation the central evaluation dimension through the integration of
failure injection, utility functions, and reconfiguration tactics.

SPEC Cloud, focusing on the infrastructure level, defines benchmarking for
elasticity, scalability and provisioning time in Infrastructure-as-a-Service
environments. While SPEC Cloud's scope is cloud platforms (moving it closer to
Adaptable TeaStore), the benchmark captures elasticity behaviour and adaptation
at the resource level rather than at the application one.

More recently, adaptation benchmarks have been proposed for the context of
serverless computing~\cite{JSSTKePSCKYGPSP19}. FaaSdom and BeFaaS offer
benchmarks for modular workloads and application-centric cases to study
elasticity and cost-performance trade-offs in serverless environments, while
SeBS defines a cross-provider benchmark suite for serverless platforms,
evaluating performance, efficiency, scalability, and reliability.

We envision future versions of Adaptable TeaStore to draw inspiration from these
work. For example, it could integrate utility functions, failure injection, and
reconfiguration tactics, as present in mRUBiS, encompass elasticity metrics and
cost-awareness from SPEC Cloud, and consider workflow resilience, typical of
serverless benchmarks.

Another direction to consider in future versions of Adaptable TeaStore is that
of energy-awareness. Drawing inspiration from recent work on energy-aware
self-adaptive systems~\cite{DMCBC25,GABDFGPPSVZ25}, the architecture could
integrate power meters to monitor energy consumption at the service level,
enabling fine-grained visibility into the energy footprint of individual
components. The architecture would then support reconfigurations that drive the
switching among service flavours based on their energy profiles during different
workload patterns. In general, the integrating of energy profiling of service
flavours with runtime adaptation mechanisms would enable researchers to
systematically explore the trade-offs between energy efficiency, performance,
and availability in cloud-native architectures.

\section{Conclusion}
\label{sec:conclusion}

In this work, we introduced Adaptable TeaStore, a novel extension of the widely
adopted TeaStore reference architecture that introduces adaptation as a
first-class concern. Unlike existing benchmarks, which primarily target
performance and scalability, Adaptable TeaStore defines explicit scenarios
regarding reconfiguration, graceful degradation, and provider substitution.
Through the introduction of mandatory and optional services, outsourceable
dependencies, multiple service flavours, and local fallback caches, our proposal
supports the systematic exploration of adaptation mechanisms under realistic,
diverse operating conditions.

The considered adaptation scenarios, ranging cyberattacks, cloud provider
outages, benign and malicious traffic surges, and DevOps-driven
reconfigurations, constitute a foundation for evaluating the effectiveness,
robustness, and trade-offs of adaptive strategies across heterogeneous cloud
environments. In this way, Adaptable TeaStore bridges a long-standing gap
between performance-oriented benchmarks and adaptation-centric ones.

Looking ahead, we see Adaptable TeaStore as a \emph{community benchmark} that
can evolve with emerging challenges in adaptive cloud systems. Several research
directions naturally follow. For instance, the integration of explicit utility
functions, failure injection, and reconfiguration tactics to enable principled
trade-off analyses between availability, cost, and quality of service. Moreover,
one can extend the architecture to include other resource-aware adaptation
policies, such as energy, advancing the study of sustainable computing. In
addition, hybridising the architecture to encompass serverless workloads,
workflow-driven services, and cross-provider federation would make Adaptable
TeaStore as an even more suitable and comprehensive testbed for the next
generation of cloud-native systems.

We envision that Adaptable TeaStore can, over time, become a reference point for
the evaluation of adaptive cloud systems. While it is not intended as a
definitive solution, its role as a common testbed may help consolidate empirical
practices, inspired and evaluate theoretical approaches, and support a more
systematic comparison of adaptation techniques. In this sense, Adaptable
TeaStore contributes to the longer-term goal of grounding self-adaptive software
research in reproducible and practically relevant experimentation.

We see this effort as a community-driven one, where members can contribute
scenarios, extensions, and evaluation methodologies, turning Adaptable TeaStore
into a shared experimental ground for reproducibility, comparability, and
innovation. The long-term aim is to establish an evolving benchmark that not
only reflects today's needs but also anticipates tomorrow's demands in the
engineering of self-adaptive software architectures. 
\bibliographystyle{eptcs}
\bibliography{biblio}

\begin{thebibliography}{10}
\providecommand{\bibitemdeclare}[2]{}
\providecommand{\surnamestart}{}
\providecommand{\surnameend}{}
\providecommand{\urlprefix}{Available at }
\providecommand{\url}[1]{\texttt{#1}}
\providecommand{\href}[2]{\texttt{#2}}
\providecommand{\urlalt}[2]{\href{#1}{#2}}
\providecommand{\doi}[1]{doi:\urlalt{https://doi.org/#1}{#1}}
\providecommand{\eprint}[1]{arXiv:\urlalt{https://arxiv.org/abs/#1}{#1}}
\providecommand{\bibinfo}[2]{#2}

\bibitemdeclare{misc}{acmeair}
\bibitem{acmeair}
\bibinfo{author}{\surnamestart {Acme Air Project}\surnameend}
  (\bibinfo{year}{2015}): \emph{\bibinfo{title}{{Acme Air} Sample and
  Benchmark}}.
\newblock \bibinfo{howpublished}{\url{https://github.com/acmeair/acmeair}}.
\newblock \bibinfo{note}{Accessed September 2025}.

\bibitemdeclare{inproceedings}{BSW17}
\bibitem{BSW17}
\bibinfo{author}{Salman \surnamestart Baset\surnameend},
  \bibinfo{author}{Marcio \surnamestart Silva\surnameend} \&
  \bibinfo{author}{Nicholas \surnamestart Wakou\surnameend}
  (\bibinfo{year}{2017}): \emph{\bibinfo{title}{{SPEC} Cloud{\texttrademark}
  {IaaS} 2016 Benchmark}}.
\newblock In \bibinfo{editor}{Walter \surnamestart Binder\surnameend},
  \bibinfo{editor}{Vittorio \surnamestart Cortellessa\surnameend},
  \bibinfo{editor}{Anne \surnamestart Koziolek\surnameend},
  \bibinfo{editor}{Evgenia \surnamestart Smirni\surnameend} \&
  \bibinfo{editor}{Meikel \surnamestart Poess\surnameend}, editors: {\slshape
  \bibinfo{booktitle}{Proceedings of the 8th {ACM/SPEC} on International
  Conference on Performance Engineering, {ICPE} 2017, L'Aquila, Italy, April
  22-26, 2017}}, \bibinfo{publisher}{{ACM}}, p. \bibinfo{pages}{423},
  \doi{10.1145/3030207.3053675}.

\bibitemdeclare{inproceedings}{CKBPH21}
\bibitem{CKBPH21}
\bibinfo{author}{Marcin \surnamestart Copik\surnameend},
  \bibinfo{author}{Grzegorz \surnamestart Kwasniewski\surnameend},
  \bibinfo{author}{Maciej \surnamestart Besta\surnameend},
  \bibinfo{author}{Michal \surnamestart Podstawski\surnameend} \&
  \bibinfo{author}{Torsten \surnamestart Hoefler\surnameend}
  (\bibinfo{year}{2021}): \emph{\bibinfo{title}{{SeBS}: a serverless benchmark
  suite for function-as-a-service computing}}.
\newblock In \bibinfo{editor}{Kaiwen \surnamestart Zhang\surnameend},
  \bibinfo{editor}{Abdelouahed \surnamestart Gherbi\surnameend},
  \bibinfo{editor}{Nalini \surnamestart Venkatasubramanian\surnameend} \&
  \bibinfo{editor}{Lu{\'{\i}}s \surnamestart Veiga\surnameend}, editors:
  {\slshape \bibinfo{booktitle}{Middleware '21: 22nd International Middleware
  Conference, Qu{\'{e}}bec City, Canada, December 6 - 10, 2021}},
  \bibinfo{publisher}{{ACM}}, pp. \bibinfo{pages}{64--78},
  \doi{10.1145/3464298.3476133}.

\bibitemdeclare{inproceedings}{DMCBC25}
\bibitem{DMCBC25}
\bibinfo{author}{Henrique \surnamestart {De Medeiros}\surnameend},
  \bibinfo{author}{Denisse \surnamestart Mu{\~{n}}ante\surnameend},
  \bibinfo{author}{Sophie \surnamestart Chabridon\surnameend},
  \bibinfo{author}{{C{\'{e}}sar Perdig{\~{a}}o} \surnamestart
  Batista\surnameend} \& \bibinfo{author}{Denis \surnamestart Conan\surnameend}
  (\bibinfo{year}{2025}): \emph{\bibinfo{title}{Adaptable {TeaStore} with
  Energy Consumption Awareness: A Case Study}}.
\newblock In \bibinfo{editor}{Giuseppe \surnamestart {De Palma}\surnameend} \&
  \bibinfo{editor}{Saverio \surnamestart Giallorenzo\surnameend}, editors:
  {\slshape \bibinfo{booktitle}{Post-proceedings of the Workshop on Adaptable
  Cloud Architectures (WACA 2025)}}, \bibinfo{series}{\thisvolume{7}}.

\bibitemdeclare{incollection}{DGLMMMS17}
\bibitem{DGLMMMS17}
\bibinfo{author}{Nicola \surnamestart Dragoni\surnameend},
  \bibinfo{author}{Saverio \surnamestart Giallorenzo\surnameend},
  \bibinfo{author}{Alberto \surnamestart Lluch{-}Lafuente\surnameend},
  \bibinfo{author}{Manuel \surnamestart Mazzara\surnameend},
  \bibinfo{author}{Fabrizio \surnamestart Montesi\surnameend},
  \bibinfo{author}{Ruslan \surnamestart Mustafin\surnameend} \&
  \bibinfo{author}{Larisa \surnamestart Safina\surnameend}
  (\bibinfo{year}{2017}): \emph{\bibinfo{title}{Microservices: Yesterday,
  Today, and Tomorrow}}.
\newblock In: {\slshape \bibinfo{booktitle}{Present and Ulterior Software
  Engineering}}, \bibinfo{publisher}{Springer}, pp. \bibinfo{pages}{195--216},
  \doi{10.1007/978-3-319-67425-4\_12}.

\bibitemdeclare{inproceedings}{GZCSRKBHRJHPH19}
\bibitem{GZCSRKBHRJHPH19}
\bibinfo{author}{Yu~\surnamestart Gan\surnameend}, \bibinfo{author}{Yanqi
  \surnamestart Zhang\surnameend}, \bibinfo{author}{Dailun \surnamestart
  Cheng\surnameend}, \bibinfo{author}{Ankitha \surnamestart Shetty\surnameend},
  \bibinfo{author}{Priyal \surnamestart Rathi\surnameend},
  \bibinfo{author}{Nayan \surnamestart Katarki\surnameend},
  \bibinfo{author}{Ariana \surnamestart Bruno\surnameend},
  \bibinfo{author}{Justin \surnamestart Hu\surnameend}, \bibinfo{author}{Brian
  \surnamestart Ritchken\surnameend}, \bibinfo{author}{Brendon \surnamestart
  Jackson\surnameend}, \bibinfo{author}{Kelvin \surnamestart Hu\surnameend},
  \bibinfo{author}{Meghna \surnamestart Pancholi\surnameend},
  \bibinfo{author}{Yuan \surnamestart He\surnameend}, \bibinfo{author}{Brett
  \surnamestart Clancy\surnameend}, \bibinfo{author}{Chris \surnamestart
  Colen\surnameend}, \bibinfo{author}{Fukang \surnamestart Wen\surnameend},
  \bibinfo{author}{Catherine \surnamestart Leung\surnameend},
  \bibinfo{author}{Siyuan \surnamestart Wang\surnameend}, \bibinfo{author}{Leon
  \surnamestart Zaruvinsky\surnameend}, \bibinfo{author}{Mateo \surnamestart
  Espinosa\surnameend}, \bibinfo{author}{Rick \surnamestart Lin\surnameend},
  \bibinfo{author}{Zhongling \surnamestart Liu\surnameend},
  \bibinfo{author}{Jake \surnamestart Padilla\surnameend} \&
  \bibinfo{author}{Christina \surnamestart Delimitrou\surnameend}
  (\bibinfo{year}{2019}): \emph{\bibinfo{title}{An Open-Source Benchmark Suite
  for Microservices and Their Hardware-Software Implications for Cloud {\&}
  Edge Systems}}.
\newblock In \bibinfo{editor}{Iris \surnamestart Bahar\surnameend},
  \bibinfo{editor}{Maurice \surnamestart Herlihy\surnameend},
  \bibinfo{editor}{Emmett \surnamestart Witchel\surnameend} \&
  \bibinfo{editor}{Alvin~R. \surnamestart Lebeck\surnameend}, editors:
  {\slshape \bibinfo{booktitle}{Proceedings of the Twenty-Fourth International
  Conference on Architectural Support for Programming Languages and Operating
  Systems, {ASPLOS} 2019, Providence, RI, USA, April 13-17, 2019}},
  \bibinfo{publisher}{{ACM}}, pp. \bibinfo{pages}{3--18},
  \doi{10.1145/3297858.3304013}.

\bibitemdeclare{article}{GABDFGPPSVZ25}
\bibitem{GABDFGPPSVZ25}
\bibinfo{author}{Simone \surnamestart Gazza\surnameend},
  \bibinfo{author}{Roberto \surnamestart Amadini\surnameend},
  \bibinfo{author}{Antonio \surnamestart Brogi\surnameend},
  \bibinfo{author}{Andrea \surnamestart D'Iapico\surnameend},
  \bibinfo{author}{Stefano \surnamestart Forti\surnameend},
  \bibinfo{author}{Saverio \surnamestart Giallorenzo\surnameend},
  \bibinfo{author}{Pierluigi \surnamestart Plebani\surnameend},
  \bibinfo{author}{Francisco \surnamestart Ponce\surnameend},
  \bibinfo{author}{Jacopo \surnamestart Soldani\surnameend},
  \bibinfo{author}{Monica \surnamestart Vitali\surnameend} \&
  \bibinfo{author}{Gianluigi \surnamestart Zavattaro\surnameend}
  (\bibinfo{year}{2025}): \emph{\bibinfo{title}{A Constraint-Based Approach to
  Optimise {QoS}- and Energy-Aware Cloud-Edge Application Deployments}}.
\newblock {\slshape \bibinfo{journal}{ACM Trans. Internet Technol.}},
  \doi{10.1145/3757061}.

\bibitemdeclare{inproceedings}{GPBSZB21}
\bibitem{GPBSZB21}
\bibinfo{author}{Martin \surnamestart Grambow\surnameend},
  \bibinfo{author}{Tobias \surnamestart Pfandzelter\surnameend},
  \bibinfo{author}{Luk \surnamestart Burchard\surnameend},
  \bibinfo{author}{Carsten \surnamestart Schubert\surnameend},
  \bibinfo{author}{Max~Xiaohang \surnamestart Zhao\surnameend} \&
  \bibinfo{author}{David \surnamestart Bermbach\surnameend}
  (\bibinfo{year}{2021}): \emph{\bibinfo{title}{{BeFaaS}: An
  Application-Centric Benchmarking Framework for {FaaS} Platforms}}.
\newblock In: {\slshape \bibinfo{booktitle}{{IEEE} International Conference on
  Cloud Engineering, {IC2E} 2021, San Francisco, CA, USA, October 4-8, 2021}},
  \bibinfo{publisher}{{IEEE}}, pp. \bibinfo{pages}{1--8},
  \doi{10.1109/IC2E52221.2021.00014}.

\bibitemdeclare{inproceedings}{GCFMIS17}
\bibitem{GCFMIS17}
\bibinfo{author}{Giona \surnamestart Granchelli\surnameend},
  \bibinfo{author}{Mario \surnamestart Cardarelli\surnameend},
  \bibinfo{author}{Paolo~Di \surnamestart Francesco\surnameend},
  \bibinfo{author}{Ivano \surnamestart Malavolta\surnameend},
  \bibinfo{author}{Ludovico \surnamestart Iovino\surnameend} \&
  \bibinfo{author}{Amleto~Di \surnamestart Salle\surnameend}
  (\bibinfo{year}{2017}): \emph{\bibinfo{title}{{MicroART}: {A} Software
  Architecture Recovery Tool for Maintaining Microservice-Based Systems}}.
\newblock In: {\slshape \bibinfo{booktitle}{2017 {IEEE} International
  Conference on Software Architecture Workshops, {ICSA} Workshops 2017,
  Gothenburg, Sweden, April 5-7, 2017}}, \bibinfo{publisher}{{IEEE} Computer
  Society}, pp. \bibinfo{pages}{298--302}, \doi{10.1109/ICSAW.2017.9}.

\bibitemdeclare{article}{JSSTKPSCKYGPSP19}
\bibitem{JSSTKPSCKYGPSP19}
\bibinfo{author}{Eric \surnamestart Jonas\surnameend}, \bibinfo{author}{Johann
  \surnamestart Schleier{-}Smith\surnameend}, \bibinfo{author}{Vikram
  \surnamestart Sreekanti\surnameend}, \bibinfo{author}{Chia{-}Che
  \surnamestart Tsai\surnameend}, \bibinfo{author}{Anurag \surnamestart
  Khandelwal\surnameend}, \bibinfo{author}{Qifan \surnamestart Pu\surnameend},
  \bibinfo{author}{Vaishaal \surnamestart Shankar\surnameend},
  \bibinfo{author}{Jo{\~{a}}o \surnamestart Carreira\surnameend},
  \bibinfo{author}{Karl \surnamestart Krauth\surnameend},
  \bibinfo{author}{Neeraja~Jayant \surnamestart Yadwadkar\surnameend},
  \bibinfo{author}{Joseph~E. \surnamestart Gonzalez\surnameend},
  \bibinfo{author}{Raluca~Ada \surnamestart Popa\surnameend},
  \bibinfo{author}{Ion \surnamestart Stoica\surnameend} \&
  \bibinfo{author}{David~A. \surnamestart Patterson\surnameend}
  (\bibinfo{year}{2019}): \emph{\bibinfo{title}{Cloud Programming Simplified:
  {A} Berkeley View on Serverless Computing}}.
\newblock {\slshape \bibinfo{journal}{CoRR}} \bibinfo{volume}{abs/1902.03383}.
\newblock \eprint{1902.03383}.

\bibitemdeclare{article}{JSSTKePSCKYGPSP19}
\bibitem{JSSTKePSCKYGPSP19}
\bibinfo{author}{Eric \surnamestart Jonas\surnameend}, \bibinfo{author}{Johann
  \surnamestart Schleier{-}Smith\surnameend}, \bibinfo{author}{Vikram
  \surnamestart Sreekanti\surnameend}, \bibinfo{author}{Chia{-}Che
  \surnamestart Tsai\surnameend}, \bibinfo{author}{Anurag \surnamestart
  Khandelwal\surnameend}, \bibinfo{author}{Qifan \surnamestart Pu\surnameend},
  \bibinfo{author}{Vaishaal \surnamestart Shankar\surnameend},
  \bibinfo{author}{Jo{\~{a}}o \surnamestart Carreira\surnameend},
  \bibinfo{author}{Karl \surnamestart Krauth\surnameend},
  \bibinfo{author}{Neeraja~Jayant \surnamestart Yadwadkar\surnameend},
  \bibinfo{author}{Joseph~E. \surnamestart Gonzalez\surnameend},
  \bibinfo{author}{Raluca~Ada \surnamestart Popa\surnameend},
  \bibinfo{author}{Ion \surnamestart Stoica\surnameend} \&
  \bibinfo{author}{David~A. \surnamestart Patterson\surnameend}
  (\bibinfo{year}{2019}): \emph{\bibinfo{title}{Cloud Programming Simplified:
  {A} Berkeley View on Serverless Computing}}.
\newblock {\slshape \bibinfo{journal}{CoRR}} \bibinfo{volume}{abs/1902.03383}.
\newblock \eprint{1902.03383}.

\bibitemdeclare{inproceedings}{KESBGK18}
\bibitem{KESBGK18}
\bibinfo{author}{J{\'{o}}akim \surnamestart von Kistowski\surnameend},
  \bibinfo{author}{Simon \surnamestart Eismann\surnameend},
  \bibinfo{author}{Norbert \surnamestart Schmitt\surnameend},
  \bibinfo{author}{Andr{\'{e}} \surnamestart Bauer\surnameend},
  \bibinfo{author}{Johannes \surnamestart Grohmann\surnameend} \&
  \bibinfo{author}{Samuel \surnamestart Kounev\surnameend}
  (\bibinfo{year}{2018}): \emph{\bibinfo{title}{{TeaStore}: {A} Micro-Service
  Reference Application for Benchmarking, Modeling and Resource Management
  Research}}.
\newblock In: {\slshape \bibinfo{booktitle}{26th {IEEE} International Symposium
  on Modeling, Analysis, and Simulation of Computer and Telecommunication
  Systems, {MASCOTS} 2018, Milwaukee, WI, USA, September 25-28, 2018}},
  \bibinfo{publisher}{{IEEE} Computer Society}, pp. \bibinfo{pages}{223--236},
  \doi{10.1109/MASCOTS.2018.00030}.

\bibitemdeclare{inproceedings}{MFKS20}
\bibitem{MFKS20}
\bibinfo{author}{Pascal \surnamestart Maissen\surnameend},
  \bibinfo{author}{Pascal \surnamestart Felber\surnameend},
  \bibinfo{author}{Peter~G. \surnamestart Kropf\surnameend} \&
  \bibinfo{author}{Valerio \surnamestart Schiavoni\surnameend}
  (\bibinfo{year}{2020}): \emph{\bibinfo{title}{{FaaSdom}: a benchmark suite
  for serverless computing}}.
\newblock In \bibinfo{editor}{Julien \surnamestart Gascon{-}Samson\surnameend},
  \bibinfo{editor}{Kaiwen \surnamestart Zhang\surnameend},
  \bibinfo{editor}{Khuzaima \surnamestart Daudjee\surnameend} \&
  \bibinfo{editor}{Bettina \surnamestart Kemme\surnameend}, editors: {\slshape
  \bibinfo{booktitle}{14th {ACM} International Conference on Distributed and
  Event-based Systems, {DEBS} 2020, Montreal, Quebec, Canada, July 13-17,
  2020}}, \bibinfo{publisher}{{ACM}}, pp. \bibinfo{pages}{73--84},
  \doi{10.1145/3401025.3401738}.

\bibitemdeclare{misc}{TeaStore-repo}
\bibitem{TeaStore-repo}
\emph{\bibinfo{title}{{TeaStore} repository}}.
\newblock
  \bibinfo{howpublished}{\url{https://github.com/DescartesResearch/TeaStore}}.
\newblock \bibinfo{note}{Accessed September 2025}.

\bibitemdeclare{inproceedings}{UNO16}
\bibitem{UNO16}
\bibinfo{author}{Takanori \surnamestart Ueda\surnameend},
  \bibinfo{author}{Takuya \surnamestart Nakaike\surnameend} \&
  \bibinfo{author}{Moriyoshi \surnamestart Ohara\surnameend}
  (\bibinfo{year}{2016}): \emph{\bibinfo{title}{Workload characterization for
  microservices}}.
\newblock In: {\slshape \bibinfo{booktitle}{2016 {IEEE} International Symposium
  on Workload Characterization, {IISWC} 2016, Providence, RI, USA, September
  25-27, 2016}}, \bibinfo{publisher}{{IEEE} Computer Society}, pp.
  \bibinfo{pages}{85--94}, \doi{10.1109/IISWC.2016.7581269}.

\bibitemdeclare{inproceedings}{V18}
\bibitem{V18}
\bibinfo{author}{Thomas \surnamestart Vogel\surnameend} (\bibinfo{year}{2018}):
  \emph{\bibinfo{title}{{mRUBiS}: an exemplar for model-based architectural
  self-healing and self-optimization}}.
\newblock In \bibinfo{editor}{Jesper \surnamestart Andersson\surnameend} \&
  \bibinfo{editor}{Danny \surnamestart Weyns\surnameend}, editors: {\slshape
  \bibinfo{booktitle}{Proceedings of the 13th International Conference on
  Software Engineering for Adaptive and Self-Managing Systems, SEAMS\@ICSE
  2018, Gothenburg, Sweden, May 28-29, 2018}}, \bibinfo{publisher}{{ACM}}, pp.
  \bibinfo{pages}{101--107}, \doi{10.1145/3194133.3194161}.

\bibitemdeclare{misc}{sockshop}
\bibitem{sockshop}
\bibinfo{author}{\surnamestart Weaveworks\surnameend} (\bibinfo{year}{2017}):
  \emph{\bibinfo{title}{Sock Shop: A Microservice Demo Application}}.
\newblock
  \bibinfo{howpublished}{\url{https://www.infoq.com/articles/sock-shop/}}.
\newblock \bibinfo{note}{Accessed September 2025}.

\bibitemdeclare{inproceedings}{ZBQ25}
\bibitem{ZBQ25}
\bibinfo{author}{Brice~Arl{\'{e}}on \surnamestart {Zemtsop Ndadji}\surnameend},
  \bibinfo{author}{Simon \surnamestart Bliudze\surnameend} \&
  \bibinfo{author}{Cl\'ement \surnamestart Quinton\surnameend}
  (\bibinfo{year}{2025}): \emph{\bibinfo{title}{{AdaptiFlow}: An Extensible
  Framework for Event-Driven Autonomy in Cloud Microservices}}.
\newblock In \bibinfo{editor}{Giuseppe \surnamestart {De Palma}\surnameend} \&
  \bibinfo{editor}{Saverio \surnamestart Giallorenzo\surnameend}, editors:
  {\slshape \bibinfo{booktitle}{Post-proceedings of the Workshop on Adaptable
  Cloud Architectures (WACA 2025)}}, \bibinfo{series}{\thisvolume{8}}.

\bibitemdeclare{inproceedings}{ZPXSXJZ18}
\bibitem{ZPXSXJZ18}
\bibinfo{author}{Xiang \surnamestart Zhou\surnameend}, \bibinfo{author}{Xin
  \surnamestart Peng\surnameend}, \bibinfo{author}{Tao \surnamestart
  Xie\surnameend}, \bibinfo{author}{Jun \surnamestart Sun\surnameend},
  \bibinfo{author}{Chenjie \surnamestart Xu\surnameend}, \bibinfo{author}{Chao
  \surnamestart Ji\surnameend} \& \bibinfo{author}{Wenyun \surnamestart
  Zhao\surnameend} (\bibinfo{year}{2018}): \emph{\bibinfo{title}{Benchmarking
  microservice systems for software engineering research}}.
\newblock In \bibinfo{editor}{Michel \surnamestart Chaudron\surnameend},
  \bibinfo{editor}{Ivica \surnamestart Crnkovic\surnameend},
  \bibinfo{editor}{Marsha \surnamestart Chechik\surnameend} \&
  \bibinfo{editor}{Mark \surnamestart Harman\surnameend}, editors: {\slshape
  \bibinfo{booktitle}{Proceedings of the 40th International Conference on
  Software Engineering: Companion Proceeedings, {ICSE} 2018, Gothenburg,
  Sweden, May 27 - June 03, 2018}}, \bibinfo{publisher}{{ACM}}, pp.
  \bibinfo{pages}{323--324}, \doi{10.1145/3183440.3194991}.

\end{thebibliography}

\end{document}